\documentclass[lettersize,journal]{IEEEtran}
\usepackage{amsmath,amsfonts}
\usepackage{algorithmic}
\usepackage{array}
\usepackage[caption=false,font=normalsize,labelfont=sf,textfont=sf]{subfig}
\usepackage{textcomp}
\usepackage{stfloats}
\usepackage{url}
\usepackage{verbatim}
\usepackage{graphicx}
\usepackage{cuted} 
\def\BibTeX{{\rm B\kern-.05em{\sc i\kern-.025em b}\kern-.08em
    T\kern-.1667em\lower.7ex\hbox{E}\kern-.125emX}}
\usepackage{balance}
\usepackage{afterpage}

\begin{document}
\title{Transient Stability Analysis of Grid-Forming Converters with Current Limiting Considering Asymmetrical Grid Faults}
\author{Seongyeon Kim, Ki-Hyun Kim, \textit{Graduate Student Member}, \textit{IEEE},\\Shenghui Cui, and Jae-Jung Jung, \textit{Member}, \textit{IEEE}}

\markboth{Transactions on Power Electronics}%
{How to Use the IEEEtran \LaTeX \ Templates}

\maketitle

\begin{abstract}
Under asymmetrical faults, analyzing the transient stability of grid-forming voltage-source converters (GFM-VSCs) becomes essential because their behavior fundamentally differs from that under symmetrical faults.
When current limiting is activated under asymmetrical faults, the point-of-common-coupling voltage of a GFM-VSC contains both positive- and negative-sequence components, and the interaction between these components generates a non-negligible negative-sequence-driven active power.
However, the transient stability of GFM-VSCs under asymmetrical faults has not been sufficiently investigated, and the influence of negative-sequence-driven active power remains unclear.
Accordingly, this letter derives the $P$–$\delta$ curve of a GFM-VSC with an elliptical current limiter under asymmetrical faults by explicitly accounting for negative-sequence effects.
This enables a more accurate transient stability assessment when extending conventional symmetrical-fault analyses to asymmetrical conditions.
The theoretical analysis is validated by the agreement between the derived $P$–$\delta$ curve and both simulation and experimental results.

\end{abstract}


\section{Introduction}
\IEEEPARstart{I}{n} recent years, research on grid-forming voltage-source converters (GFM-VSCs) has predominantly focused on symmetrical fault conditions.
In practical power systems, however, asymmetrical faults occur more frequently; thus, understanding the behavior of GFM-VSCs under such conditions is essential \cite{Asymmetric_many}.

Since a GFM-VSC is modeled as a positive-sequence voltage source behind a series impedance, it is inherently vulnerable to overcurrent during grid faults.
This vulnerability arises because the voltage-source control structure attempts to maintain the positive-sequence voltage while forcing the negative-sequence voltage toward zero, thereby naturally driving excessive fault currents.
Therefore, implementing an overcurrent limiting algorithm is indispensable \cite{Fan_review}.
Moreover, unlike the symmetrical fault case, in which limiting only the positive-sequence current suffices to keep all phase currents within allowable limits, the asymmetrical fault case requires considering both positive- and negative-sequence currents \cite{Baeckeland_review}.
As a result, the transient stability characteristics of a GFM-VSC with current limiting differ fundamentally between symmetrical and asymmetrical faults, necessitating a dedicated analysis.

Under asymmetrical faults, the electromotive force (EMF) of a synchronous generator remains purely positive-sequence, even though negative-sequence currents may flow.
In contrast, a GFM-VSC with an overcurrent limiting algorithm synthesizes a nonzero negative-sequence point-of-common-coupling (PCC) voltage in order to regulate the negative-sequence current within safe operating limits.
Consequently, the interaction between the negative-sequence PCC voltage and current produces a non-negligible negative-sequence–driven active power—an effect absent in synchronous generators because the negative-sequence component of their EMF remains zero.
This additional power component alters the power-angle characteristics and must therefore be explicitly considered in transient stability analysis.

Nevertheless, the transient stability of GFM-VSCs under asymmetrical faults has received limited attention, and the impact of negative-sequence–driven active power has largely remained unexplained \cite{New_p-delta}.
Motivated by this gap, this letter analyzes the transient stability of a GFM-VSC under asymmetrical faults by deriving the $P$–$\delta$ curve while explicitly incorporating the effects of both positive- and negative-sequence components and the activation of the elliptical current limiter (ECL).
This formulation allows the angle-dynamics analysis traditionally used for symmetrical faults to be extended to asymmetrical cases.
The theoretical analysis is validated by the consistency among the derived $P$–$\delta$ curve, simulation results, and experimental results
\begin{figure}[!t]
\centering
\includegraphics[width=0.45\textwidth]{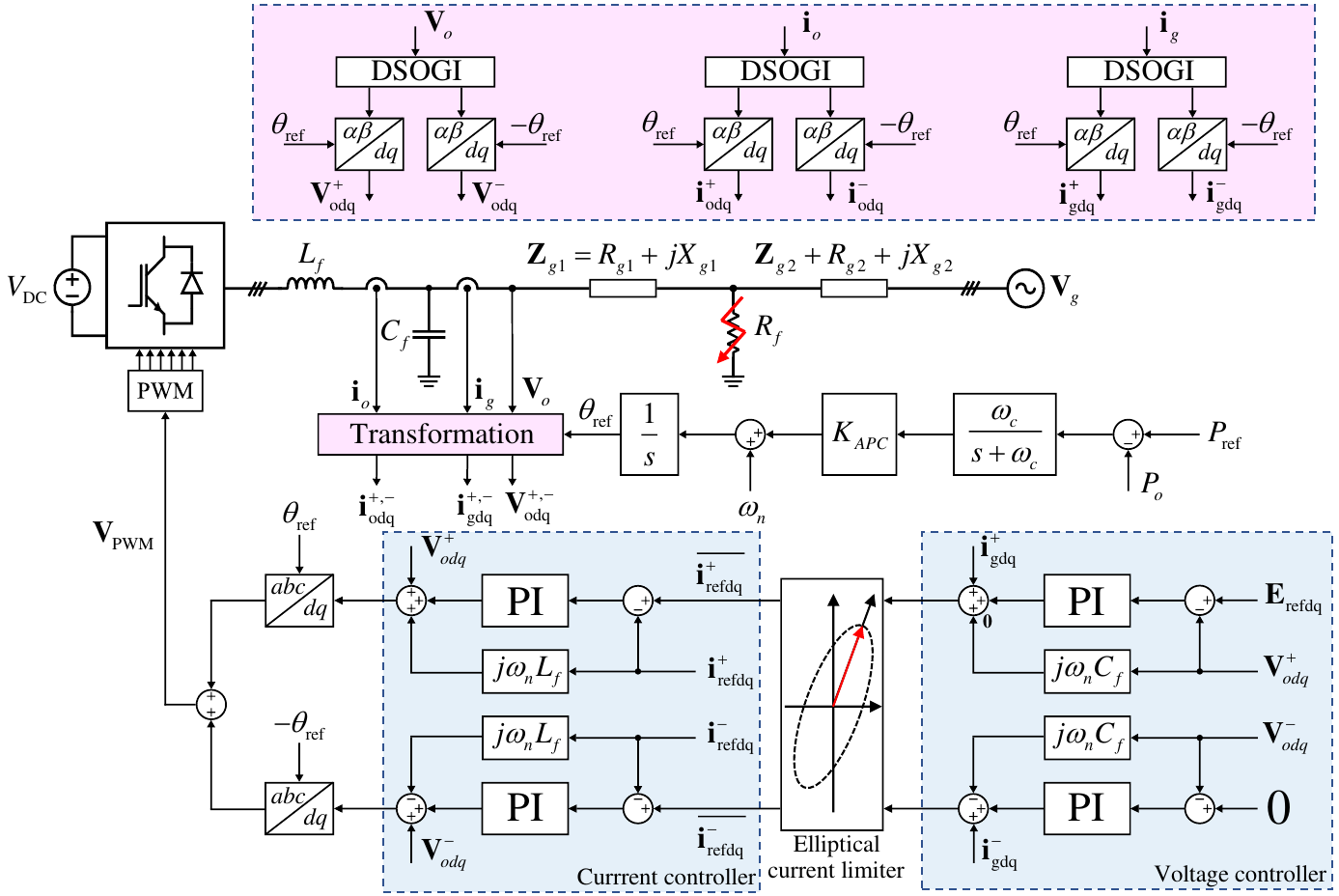}
\caption{Block diagram of a GFM-VSC with an elliptical current limiter.}
\label{fig1}
\end{figure}

\section{System Description}
Fig. 1 illustrates the block diagram of a GFM-VSC with an ECL.
$V_{\mathrm{DC}}, L_f, C_f, R_f, \mathbf{Z}_g,$ and $\mathbf{V}_g$ denote the DC-link voltage, filter inductance, filter capacitance, fault resistance, grid impedance, and grid voltage, respectively.
$\mathbf{E}_{\mathrm{ref}}, \mathbf{V}_o, \mathbf{i}_o, \mathbf{i}_g, \mathbf{i}_{\mathrm{ref}}^{\mathrm{+(-)}}, \overline{\mathbf{i}_{\mathrm{ref}}^{\mathrm{+(-)}}}$ denote the voltage reference, the PCC voltage, the converter-side current, the grid-side current, and the current references before and after the limiter, respectively.
Moreover, $P_{\mathrm{ref}}, P_o, \omega_n, \theta_{\mathrm{ref}}, K_{APC}, \omega_c$ represent the active power reference, output active power, nominal angular frequency, angle reference, proportional gain of the active power controller, and its cutoff frequency.
The superscripts ``$+$'' and ``$-$'' denote the positive and negative sequences; the subscript ``$dq$'' denotes a complex variable in the synchronous $dq$-frame.
As shown in Fig. 1, a dual second-order generalized integrator (DSOGI) is employed to decouple and independently regulate the positive- and negative-sequence components \cite{DSOGI}.

\section{Modeling and Transient Stability Analysis of GFM-VSC with Elliptical Current Limiter}

\subsection{Implementation of Elliptical Current Limiter (ECL)}
This letter considers a current limiting scheme that protects GFM-VSCs from overcurrents under both symmetrical and asymmetrical faults.
Under asymmetrical faults, since the GFM-VSC naturally generates both positive- and negative-sequence current references to maintain the positive-sequence voltage, the current references must be scaled down by jointly considering both sequences, as in \eqref{eq:current_scaling1}.
Accordingly, among the asymmetrical current limiting approaches, this letter focuses on the elliptical current limiter (ECL), proposed in \cite{ellip_proposed} and later adopted in \cite{ellip_used0}.
The ECL scales down the positive- and negative-sequence current references by the same factor $c$, as in \eqref{eq:current_scaling2} (i.e., $c^{+}=c^{-}$), ensuring that each phase current does not exceed the maximum allowable current $I_{\mathrm{lim}}$.
By decoupling the scaling factors, i.e., setting $c^{+}\neq c^{-}$, sequence priority can be assigned between the positive- and negative-sequence currents; this modification is likewise accommodated within the analytical framework of this letter.
{
\begin{equation}
\overline{\mathbf{i}_{\mathrm{refdq}}^{\mathrm{+(-)}}}
= c^{\mathrm{+(-)}} \cdot \mathbf{i}^{\mathrm{+(-)}}_{\mathrm{refdq}}
\label{eq:current_scaling1}
\end{equation}
}
{
\begin{equation}
c = c^{\mathrm{+(-)}} = \min \!\left\{ 1,\,
{I_{\mathrm{lim}}}\bigl/{i^{\mathrm{max}}_{\mathrm{ref}}}
\right\}
\label{eq:current_scaling2}
\end{equation}
}

In \eqref{eq:current_scaling2}, $i_{\mathrm{ref}}^{\mathrm{max}}$ denotes the maximum phase current reference obtained from \eqref{eq:iref_peaks} and \eqref{eq:iref_max}, where {\small\(\lvert\mathbf{i}_{\mathrm{ref}}^{+(-)}\rvert=\sqrt{(i_{\mathrm{refd}}^{+(-)})^2+(i_{\mathrm{refq}}^{+(-)})^2}\) and \(\phi_{\mathrm{ref}}^{+(-)}=\tan^{-1}\!\bigl(i_{\mathrm{refq}}^{+(-)}/i_{\mathrm{refd}}^{+(-)}\bigr)\)}.
\newcommand{\Ip}{| \mathbf{i}^{+}_\mathrm{ref} |}
\newcommand{\Imn}{| \mathbf{i}^{-}_\mathrm{ref} |}
\newcommand{\phis}{\phi^{+}_\mathrm{ref}+\phi^{-}_\mathrm{ref}}
{
\begin{equation}
\begin{aligned}
i^{\mathrm{peak}}_{\mathrm{ref,a}}
&= \sqrt{\Ip^{2}+\Imn^{2}+2\,\Ip\,\Imn\cos(\phis)},\\
i^{\mathrm{peak}}_{\mathrm{ref,b}}
&= \sqrt{\Ip^{2}+\Imn^{2}+2\,\Ip\,\Imn\cos\!\bigl(\phis-\tfrac{2\pi}{3}\bigr)},\\
i^{\mathrm{peak}}_{\mathrm{ref,c}}
&= \sqrt{\Ip^{2}+\Imn^{2}+2\,\Ip\,\Imn\cos\!\bigl(\phis+\tfrac{2\pi}{3}\bigr)}
\end{aligned}
\label{eq:iref_peaks}
\end{equation}
}
{
\begin{equation}
i^{\mathrm{max}}_{\mathrm{ref}} = 
\max \!\bigl[
i^{\mathrm{peak}}_{\mathrm{ref,a}},
i^{\mathrm{peak}}_{\mathrm{ref,b}},
i^{\mathrm{peak}}_{\mathrm{ref,c}}
\bigr]
\label{eq:iref_max}
\end{equation}
}

When the current limiter is not activated, $i_{\mathrm{ref}}^{\mathrm{max}} < I_{\mathrm{lim}}$ holds, resulting in $c = 1$, and consequently {\small$\overline{\mathbf{i}_{\mathrm{refdq}}^{\mathrm{+(-)}}} = \mathbf{i}_{\mathrm{refdq}}^{+(-)}$}.
In contrast, when the current limiter is activated, $i_{\mathrm{ref}}^{\mathrm{max}} > I_{\mathrm{lim}}$ holds, and $c$ becomes a real number satisfying $0 < c < 1$.

\subsection{Equivalent Circuit Model of GFM-VSC with ECL}
In this section, the equivalent circuit model of the GFM-VSC with the previously described ECL is derived.
To this end, as shown in Fig. 1, the windup phenomenon occurring in the voltage controller integrator when the current limiter saturates the current reference during overcurrent conditions is first considered.
To prevent the windup phenomenon that deteriorates the dynamic performance and stability of the GFM-VSC, the output of the integrator is held at zero during the overcurrent limiting period, as described in \cite{Bofan_Req}.
As a result, as illustrated in Fig. 1, the positive- and negative-sequence current references after the ECL are derived as \eqref{eq:after_limitation_pos} and \eqref{eq:after_limitation_neg}, respectively.
{
\begin{equation}
\overline{\mathbf{i}^{+}_\mathrm{refdq}}
= {c}
\Big[
K_{P}^{V}\!\big(\mathbf{E}_{\mathrm{refdq}} - \mathbf{V}^{+}_{\mathrm{odq}}\big)
+ \mathbf{i}^{+}_{\mathrm{gdq}}
+ j \omega_{g} C_{f} \mathbf{V}^{+}_{\mathrm{odq}}
\Big]
\label{eq:after_limitation_pos}
\end{equation}
\begin{equation}
\overline{\mathbf{i}^{-}_{\mathrm{refdq}}}
= {c}
\Big[
K_{P}^{V}\!\big(\mathbf{0} - \mathbf{V}^{-}_{\mathrm{odq}}\big)
+ \mathbf{i}^{-}_{\mathrm{gdq}}
- j \omega_{g} C_{f} \mathbf{V}^{-}_{\mathrm{odq}}
\Big]
\label{eq:after_limitation_neg}
\end{equation}
}

For analytical simplicity, the dynamics of the filter capacitor is neglected, resulting in \eqref{eq:cap_dynamic_pos} and \eqref{eq:cap_dynamic_neg}.
{
\begin{equation}
\mathbf{i}^{+}_{\mathrm{odq}}
= \mathbf{i}^{+}_{\mathrm{gdq}}
+ j \omega_{n} C_{f} \mathbf{V}^{+}_{\mathrm{odq}}
\label{eq:cap_dynamic_pos}
\end{equation}
\begin{equation}
\mathbf{i}^{-}_{\mathrm{odq}}
= \mathbf{i}^{-}_{\mathrm{gdq}}
- j \omega_{n} C_{f} \mathbf{V}^{-}_{\mathrm{odq}}
\label{eq:cap_dynamic_neg}
\end{equation}
}

\begin{figure}[!t]
\centering
\includegraphics[width=0.45\textwidth]{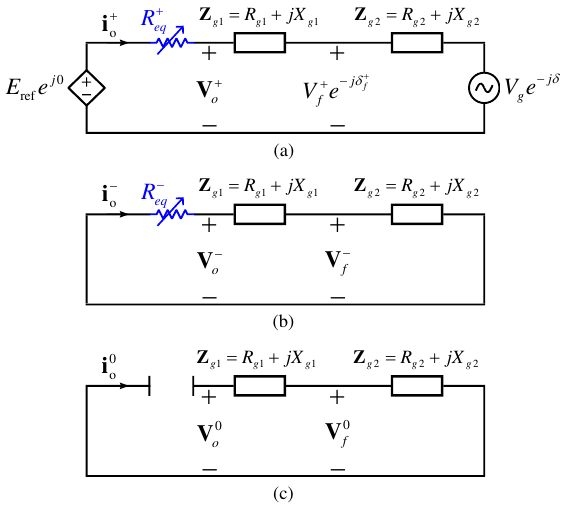}
\caption{Equivalent sequence networks of a grid-connected GFM-VSC with ECL: (a) positive-sequence. (b) negative-sequence. (c) zero-sequence.}
\label{fig2}
\end{figure}
In addition, since the bandwidth of the current controller is sufficiently wider than that of the outer controller, both the positive- and negative-sequence current controllers can be assumed to have unity gain (i.e., {\small$\overline{\mathbf{i}^{+(-)}_{\mathrm{refdq}}} = \mathbf{i}^{+(-)}_{\mathrm{odq}}$}).
Finally, by combining {\small$\overline{\mathbf{i}^{+(-)}_{\mathrm{refdq}}} = \mathbf{i}^{+(-)}_{\mathrm{odq}}$}, \eqref{eq:after_limitation_pos}--\eqref{eq:cap_dynamic_neg}, \eqref{eq:Req+} and \eqref{eq:Req-} are obtained, which confirm that, during the current-limiting period, the inner-loop controllers and the ECL in both the positive- and negative-sequence circuits are equivalently represented as resistances, as shown in Fig.~2.
This is because \(c\) is a real number satisfying \(0 < c \leq 1\).
In particular, as seen from \eqref{eq:Req+} and \eqref{eq:Req-}, $R_{eq}^{+}=R_{eq}^{-}$ holds, indicating that the positive- and negative-sequence resistances are equivalently represented with the same value during the current-limiting period.
Moreover, when the ECL is not triggered (i.e., $c=1$), {\small$\mathbf{E}_{\mathrm{refdq}}^{+}=\mathbf{V}_{\mathrm{odq}}^{+}$, $\mathbf{V}_{\mathrm{odq}}^{-}=0$, and $R_{eq}^{+}=R_{eq}^{-}=0$}, which mathematically confirms the normal condition.
{
\begin{equation}
\mathbf{E}_{\mathrm{refdq}} - \mathbf{V}^{+}_{\mathrm{odq}}
= \frac{1 - c}{c K_{P}^{V}} \mathbf{i}^{+}_{\mathrm{odq}}
\;\;\Rightarrow\;\;
R_{eq}^{+} = \frac{1 - c}{c K_{P}^{V}}
\label{eq:Req+}
\end{equation}
\begin{equation}
\mathbf{0} - \mathbf{V}^{-}_{\mathrm{odq}}
= \frac{1 - c}{c K_{P}^{V}} \mathbf{i}^{-}_{\mathrm{odq}}
\;\;\Rightarrow\;\;
R_{eq}^{-} = \frac{1 - c}{c K_{P}^{V}}
\label{eq:Req-}
\end{equation}
}

Furthermore, since a three-phase three-wire converter has no zero-sequence current path, the zero-sequence network is treated as open, as shown in Fig.~2 (c), and \(V_f\) denotes the fault-point voltage, defined later.

\subsection{Transient Stability Analysis of GFM-VSC with ECL}
In this section, the $P$--$\delta$ curve of an GFM-VSC with ECL under asymmetrical faults is derived, and its transient stability is analyzed.
As noted earlier, unlike a synchronous generator, a GFM-VSC synthesizes a negative-sequence voltage to regulate the converter current within its operating limits while accounting for the negative-sequence current.
Consequently, the negative-sequence PCC voltage of the GFM-VSC becomes nonzero, and the active power delivered to the grid under asymmetrical faults is expressed in \eqref{eq:instantaneous_power} \cite{instantaneous_power}.
{
\begin{align}
P_{o} &= \underbrace{\Re\!\big( \mathbf{V}^{+}_{o}\mathbf{i}^{+*}_{o} \big)}_{P_{o}^{+}}
+ \underbrace{\Re\!\big( \mathbf{V}^{-}_{o}\mathbf{i}^{-*}_{o} \big)}_{P_{o}^{-}}
+ \underbrace{\Re\!\big( \mathbf{V}^{+}_{o}\mathbf{i}^{-*}_{o} + \mathbf{V}^{-}_{o}\mathbf{i}^{+*}_{o} \big)}_{P_{2\omega}}
\label{eq:instantaneous_power}
\end{align}
}

As observed from \eqref{eq:instantaneous_power}, the output active power under asymmetrical fault conditions can be decomposed into the DC components, i.e., the positive-sequence active power $P_{o}^{+}$ and the negative-sequence active power $P_{o}^{-}$, and the AC component, i.e., the cross-sequence active power component $P_{2\omega}$ oscillating at twice the fundamental frequency \cite{instantaneous_power}.
The North American Electric Reliability Corporation (NERC) and National Grid ESO recommend limiting the bandwidth of the active power controller, which serves as the synchronization control loop, to below 5 Hz in order to prevent continuous sub-harmonic frequency output of the VSC and to avoid instability \cite{NERC, GBGF}.
In this letter, the bandwidth of the active power controller is also set to less than 5 Hz, and as a result, the second-harmonic component of the output active power $P_{2\omega}$ can be neglected in the transient stability analysis of the GFM-VSC.
Therefore, both \(P_{o}^{+}\) and \(P_{o}^{-}\) must be considered in the transient stability analysis of the GFM-VSC; details follow.

Under single-line-to-ground (SLG), double-line-to-ground (DLG), and line-to-line (LL) faults, the positive- and negative-sequence voltages at the fault point in Fig.~2 are given by \eqref{SLG}–\eqref{LL}, respectively \cite{Rui_liu}.
As shown in the previous section, $R_{eq}^{+}$ and $R_{eq}^{-}$ are identical; thus, they can be represented uniformly as $R_{eq}$.
\par 
\begingroup
\scriptsize
\begin{equation}
\begin{bmatrix}
\mathbf{V}_{f}^{+}\\[2pt]
\mathbf{V}_{f}^{-}
\end{bmatrix}
=
\begin{bmatrix}
\dfrac{2\mathbf{Z}_p+3R_f}{3\mathbf{Z}_p+3R_f}
\left(
  \dfrac{\mathbf{E}_{\mathrm{ref}} \mathbf{Z}_{g2}}{R_{eq}+\mathbf{Z}_{g1}+\mathbf{Z}_{g2}}
  + \dfrac{(R_{eq}+\mathbf{Z}_{g1})\,\mathbf{V}_g}{R_{eq}+\mathbf{Z}_{g1}+\mathbf{Z}_{g2}}
\right)
\\[10pt]
\dfrac{-\mathbf{Z}_p}{3\mathbf{Z}_p+3R_f}
\left(
  \dfrac{\mathbf{E}_{\mathrm{ref}} \mathbf{Z}_{g2}}{R_{eq}+\mathbf{Z}_{g1}+\mathbf{Z}_{g2}}
  + \dfrac{(R_{eq}+\mathbf{Z}_{g1})\,\mathbf{V}_g}{R_{eq}+\mathbf{Z}_{g1}+\mathbf{Z}_{g2}}
\right)
\end{bmatrix}
\label{SLG}
\end{equation}
\begin{equation}
\begin{bmatrix}
\mathbf{V}_{f}^{+}\\[2pt]
\mathbf{V}_{f}^{-}
\end{bmatrix}
=
\begin{bmatrix}
\dfrac{\mathbf{Z}_p+R_f}{2\mathbf{Z}_p+R_f}
\left(
  \dfrac{\mathbf{E}_{\mathrm{ref}} \mathbf{Z}_{g2}}{R_{eq}+\mathbf{Z}_{g1}+\mathbf{Z}_{g2}}
  + \dfrac{(R_{eq}+\mathbf{Z}_{g1})\,\mathbf{V}_g}{R_{eq}+\mathbf{Z}_{g1}+\mathbf{Z}_{g2}}
\right)
\\[10pt]
\dfrac{\mathbf{Z}_p}{2\mathbf{Z}_p+R_f}
\left(
  \dfrac{\mathbf{E}_{\mathrm{ref}} \mathbf{Z}_{g2}}{R_{eq}+\mathbf{Z}_{g1}+\mathbf{Z}_{g2}}
  + \dfrac{(R_{eq}+\mathbf{Z}_{g1})\,\mathbf{V}_g}{R_{eq}+\mathbf{Z}_{g1}+\mathbf{Z}_{g2}}
\right)
\end{bmatrix}
\label{DLG}
\end{equation}
\begin{equation}
\mathbf{V}_{f}^{+}=\mathbf{V}_{f}^{-}
= \frac{\mathbf{Z}_p+3R_f}{3\mathbf{Z}_p+6R_f}
\left(
  \frac{\mathbf{E}_{\mathrm{ref}} \mathbf{Z}_{g2}}{R_{eq}+\mathbf{Z}_{g1}+\mathbf{Z}_{g2}}
  + \dfrac{(R_{eq}+\mathbf{Z}_{g1})\,\mathbf{V}_g}{R_{eq}+\mathbf{Z}_{g1}+\mathbf{Z}_{g2}}
\right)
\label{LL}
\end{equation}
\endgroup
where \(\mathbf{Z}_p=\mathbf{Z}_n=\mathbf{Z}_{g2}(R_{eq}+\mathbf{Z}_{g1}) / (R_{eq}+\mathbf{Z}_{g1}+\mathbf{Z}_{g2})\) and \(\mathbf{Z}_0=\mathbf{Z}_{g2}\).
Utilizing \eqref{SLG}–\eqref{LL} together with Fig.~2, the output current of the GFM-VSC can be expressed as in \eqref{eq:io}.
{
\begin{equation}
\mathbf{i}_o^{+}({R_{eq}})
=
\frac{\mathbf{E}_{\mathrm{ref}}  - \mathbf{V}_f^{+}}
{R_{eq} + \mathbf{Z}_{g1}},\quad
\mathbf{i}_o^{-}({R_{eq}})
=
\frac{- \mathbf{V}_f^{-}}
{R_{eq} + \mathbf{Z}_{g1}}
\label{eq:io}
\end{equation}
}
\begin{figure}
\centering
\includegraphics[width=0.45\textwidth]{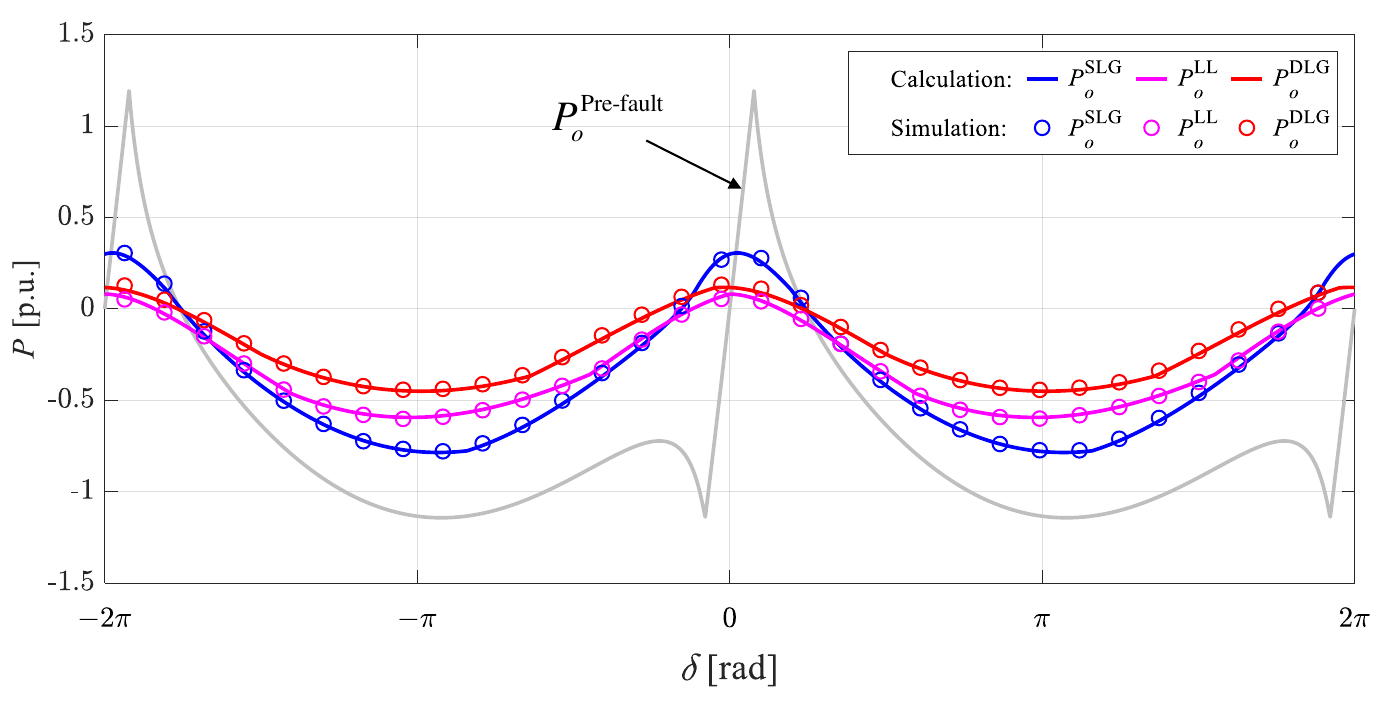}
\caption{$P\text{--}\delta$ curves under pre-fault, SLG, DLG, and LL fault conditions.}
\label{fig3}
\end{figure}
\begin{figure}[!t]
\centering
\includegraphics[width=0.45\textwidth]{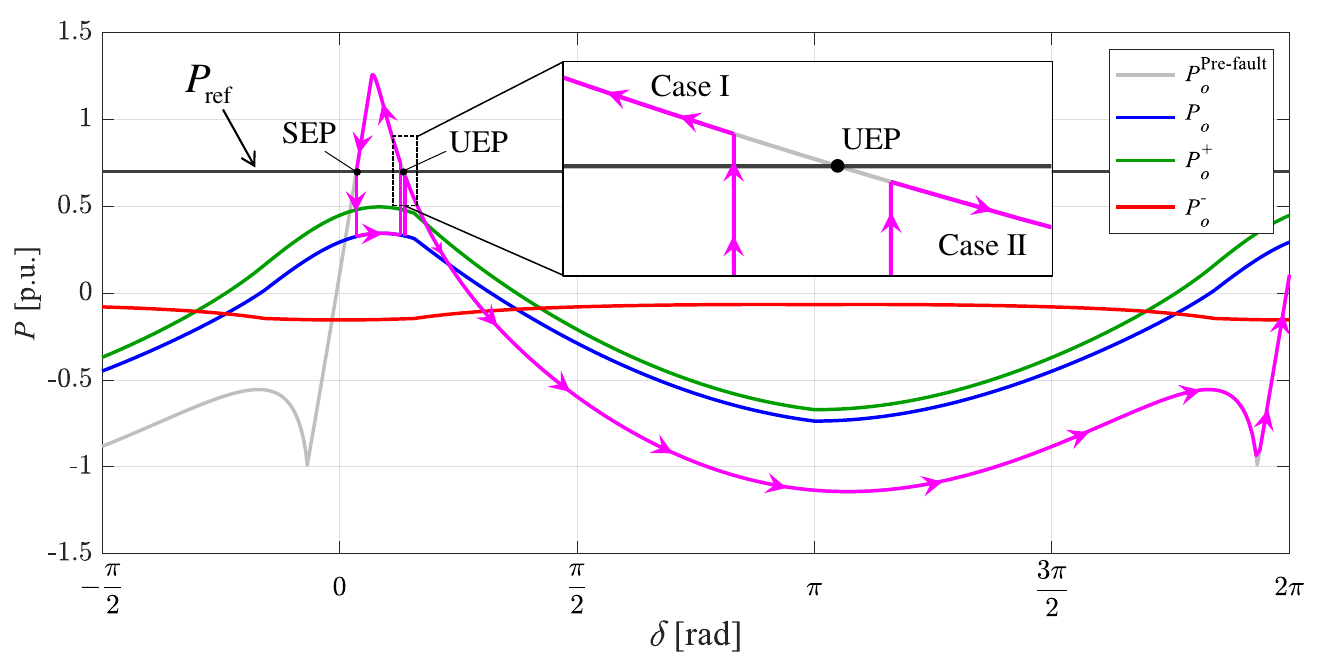}
\caption{Transient stability analysis of the GFM-VSC using \(P_o\), \(P_o^{+}\), and \(P_o^{-}\) curves under a SLG fault.}
\label{fig4}
\end{figure}

The maximum phase current \(\mathbf{i}_{o}^{\mathrm{max}} \) of the GFM-VSC can be derived by substituting \( \mathbf{i}_{\mathrm{ref}}^{+} \) and \( \mathbf{i}_{\mathrm{ref}}^{-} \) in \eqref{eq:iref_peaks}–\eqref{eq:iref_max} with \( \mathbf{i}_{\mathrm{o}}^{+} \) and \( \mathbf{i}_{\mathrm{o}}^{-} \) from \eqref{eq:io}.
Since \( \mathbf{i}_{o}^{+} \) and \( \mathbf{i}_{o}^{-} \) are both functions of \( R_{eq} \), \( \mathbf{i}_{o}^{\mathrm{max}} \) is also expressed as a function of \( R_{eq} \).
Accordingly, a solution \( R_{eq} \) can be obtained that satisfies \( \mathbf{i}_{o}^{\mathrm{max}}(R_{eq}) = I_{\mathrm{lim}} \), ensuring that, for all power angle \( \delta \), the maximum phase current reaches the allowable limit.
If \( R_{eq}=0 \) and \( \mathbf{i}_{o}^{\mathrm{max}} < I_{\mathrm{lim}} \), no current limiting is required; accordingly, \( R_{eq} \) should remain zero.
Using the \(R_{eq}\) computed for all \(\delta\) together with \eqref{eq:io} and Kirchhoff’s laws, the \( \mathbf{V}_{o}^{+} \) and \( \mathbf{V}_{o}^{-} \) shown in Fig.~2(a) and (b) can be obtained.
Finally, by applying \eqref{eq:instantaneous_power}, the \( \mathbf{P}_{o}^{+} \) and \( \mathbf{P}_{o}^{-} \) are derived as in \eqref{eq:Po_pos} and \eqref{eq:Po_neg}.
Plotting these expressions yields the $P$--$\delta$ curves shown in Fig.~3, which is consistent with EMT simulations for all three fault types (SLG, DLG, and LL).
The corresponding $P$--$\delta$ curves are obtained under the conditions $E_{\mathrm{ref}} = 1.1$ p.u., $X_{g} = 0.2$ p.u., $R_{g} = 0.04$ p.u., and $I_{\mathrm{lim}} = 1.2$ p.u..
{
\begin{equation}
\begin{aligned}
P_o^{+} (\delta) =\
&\frac{
R_{g1} \!E_{\mathrm{ref}}^{2}
+ \left(R_{eq}-R_{g1}\right) E_{\mathrm{ref}} V_{f}^{+}\cos\delta_{f}^{+}
}{
\left(R_{eq}+R_{g1}\right)^{2} + X_{g1}^{2}
}
\\[6pt]
&\quad + \frac{
E_{\mathrm{ref}} V_{f}^{+} X_{g1} \sin\delta_{f}^{+}
- R_{eq} V_{f}^{+2}
}{
\left(R_{eq}+R_{g1}\right)^{2} + X_{g1}^{2}
}
\end{aligned}
\label{eq:Po_pos}
\end{equation}
}
{
\begin{equation}
P_o^{-} (\delta) =
\frac{-\,R_{eq}\,(V_{f}^{-})^{2}}{(R_{eq}+R_{g1})^{2}+X_{g1}^{2}}
\label{eq:Po_neg}
\end{equation}}

The detailed analysis is based on the SLG fault scenario, and the resulting $P_{o}$, $P_{o}^{+}$, and $P_{o}^{-}$ curves of the GFM-VSC under the SLG condition are shown in Fig.~4.
In Fig.~4, under the pre-fault condition, the GFM-VSC operates at the stable equilibrium point (SEP), which corresponds to one of the intersections between the gray $P$--$\delta$ curve $P_{o}^{\text{Pre-fault}}$ and $P_{\mathrm{ref}}$.
When an SLG fault occurs, the $\delta$ increases along the $P_{o}$ curve.
If the fault is cleared before $\delta$ exceeds the unstable equilibrium point (UEP), the GFM-VSC can remain synchronized (Case~I).
In contrast, if the fault persists and $\delta$ exceeds the UEP, the GFM-VSC loses synchronism even after the fault is cleared (Case~II).
For qualitative analysis, the critical clearing time (CCT) is obtained by repeatedly integrating \eqref{Swing_equation} while increasing the fault duration in increments of 0.001~s.
Here, the CCT represents the maximum fault duration for which the GFM-VSC can remain stable after fault clearance.
As a result, the CCT that distinguishes Case~I from Case~II is obtained as 0.264~s, and its validation is presented in the following experiment section.
\begin{equation}
\dot{\delta}=\omega_n\omega,\quad
\dot{\omega}=\omega_c K_{APC}\!\left(P_{\mathrm{ref}}-P_o(\delta)\right)-\omega_c\omega
\label{Swing_equation}
\end{equation}
where $\omega_n= 2\pi \cdot 60~\mathrm{rad/s}$, $\omega_c= 2\pi \cdot 0.3~\mathrm{rad/s}$, $P_{\mathrm{ref}}=0.7$ p.u., and $K_{APC}=0.02$ p.u..

Furthermore, it can be observed through Fig.~4 and \eqref{eq:Po_neg} that the negative-sequence active power $P_{o}^{-}$ always remains negative.
This increases the acceleration region of the GFM-VSC under asymmetrical faults and may also lead to the disappearance of the SEP.
In other words, the $P_{o}^{-}$ always deteriorates the transient stability of the GFM-VSC under asymmetrical faults.

\begin{figure*}[!t]
\centering
\includegraphics[width=1\textwidth]{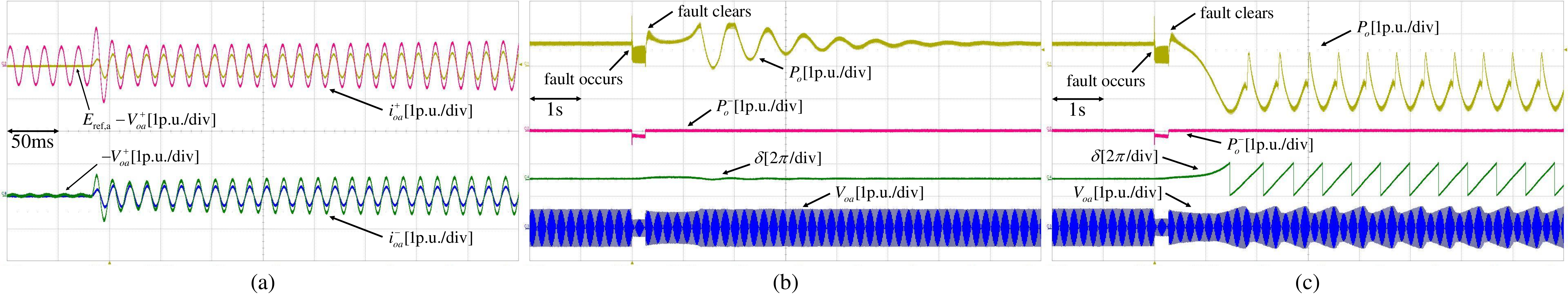}
\caption{Experimental results under an SLG fault: (a) relationships between \(E_{\mathrm{ref},a}-V_{oa}^{+}\) and \(i_{oa}^{+}\), and between \(-V_{oa}^{-}\) and \(i_{oa}^{-}\); (b) response for a 0.27~s fault duration (Case I). (c) response for a 0.28~s fault duration (Case II).}
\label{fig5}
\end{figure*}

\section{Experimental Results}
To validate the theoretical analysis, 2-kVA downscaled experiments were conducted.
The grid reactance \(X_g\) and resistance \(R_g\) were set to 0.2 p.u. and 0.04 p.u., respectively.
And the maximum allowable current \(I_{\text{lim}}\), active-power reference \(P_{\text{ref}}\), and voltage reference \(E_{\text{ref}}\) were set to 1.2 p.u., 0.7 p.u., and 1.1 p.u., respectively.
Under these conditions, an SLG fault on phase~A is experimentally applied, and the resulting $P$--$\delta$ curve of the GFM-VSC corresponds to that shown in Fig.~4.

Fig.~5(a) illustrates the relationships between $E_{\mathrm{ref,a}} - V_{oa}^{+}$ and $i_{oa}^{+}$, as well as between $-{V}_{oa}^{-}$ and $i_{oa}^{-}$ during the SLG fault.
As shown in Fig.~5(a), $E_{\mathrm{ref,a}} - V_{oa}^{+}$ and $i_{oa}^{+}$ are in phase, and likewise, $-{V}_{oa}^{-}$ and $i_{oa}^{-}$ are also in phase.
This indicates that the equivalent impedances for current limiting behave as resistances, as illustrated in Fig.~2(a) and Fig.~2(b), which is consistent with the theoretical analysis presented above.

Fig.~5(b) and Fig.~5(c) show the experimental results when the SLG fault persists for 0.27~s and 0.28~s, respectively.
When the fault lasts for 0.27~s, the GFM-VSC maintains synchronism after the fault is cleared, whereas it loses synchronism when the fault duration is 0.28~s.
This indicates that the experimentally obtained CCT lies between 0.27~s and 0.28~s, which is in close agreement with the calculated value of 0.264~s derived through repeated integration in the previous section.
Furthermore, Fig.~5(b) and Fig.~5(c) show that the negative-sequence active power $P_{o}^{-}$ is approximately $-0.15$ p.u., which is consistent with the previous analysis indicating that $P_{o}^{-}$ always remains negative.

\section{Conclusion}
This letter analyzes the transient stability of a GFM-VSC with current limiter under asymmetrical grid faults.
To this end, first, the inner-loop controller and the elliptical current limiter of the GFM-VSC are shown to be representable as identical equivalent resistances in both the positive- and negative-sequence networks.
Second, based on this finding, the $P$--$\delta$ curve under asymmetrical faults is derived.
In particular, it is shown that the negative-sequence active power is consistently negative, thereby degrading the transient stability of GFM-VSCs.


\begin{thebibliography}{1}

\bibitem{Asymmetric_many}
N. Tleis, Power Systems Modelling and Fault Analysis: Theory and Practice. Amsterdam, The Netherlands: Elsevier, 2007.

\bibitem{Fan_review}
B. Fan, T. Liu, F. Zhao, H. Wu and X. Wang, "A Review of Current-Limiting Control of Grid-Forming Inverters Under Symmetrical Disturbances," in \textit{IEEE Open J. Power Electron.}, vol. 3, pp. 955-969, 2022.

\bibitem{Baeckeland_review}
N. Baeckeland, D. Chatterjee, M. Lu, B. Johnson and G. -S. Seo, "Overcurrent Limiting in Grid-Forming Inverters: A Comprehensive Review and Discussion," in \textit{IEEE Trans. Power Electron.}, vol. 39, no. 11, pp. 14493-14517, Nov. 2024.

\bibitem{New_p-delta}
Q. Liu, M. Wang, M. Nick, C. Chen and X. Zhao, "Current-Limiting Control Design for Grid-Forming Capability Enhancement of IBRs Under Asymmetric Grid Disturbances," in \textit{IEEE Trans. Power Electron.}, early access, 14. Nov, 2024, doi: 10.1109/TPEL.2025.3632684.

\bibitem{DSOGI}
P. Rodriguez, J. Pou, J. Bergas, J. I. Candela, R. P. Burgos and D. Boroyevich, "Decoupled Double Synchronous Reference Frame PLL for Power Converters Control," in \textit{IEEE Trans. Power Electron.}, vol. 22, no. 2, pp. 584-592, March 2007.

\bibitem{ellip_proposed}
K. Schönleber, E. Prieto-Araujo, S. Ratés-Palau and O. Gomis-Bellmunt, "Extended Current Limitation for Unbalanced Faults in MMC-HVDC-Connected Wind Power Plants," in \textit{IEEE Trans. Power Deliv.}, vol. 33, no. 4, pp. 1875-1884, Aug. 2018.

\bibitem{ellip_used0}
Y. Zhang, C. Zhang, M. Molinas and X. Cai, "Control of Virtual Synchronous Generator With Improved Transient Angle Stability Under Symmetric and Asymmetric Short Circuit Fault," in \textit{IEEE Trans. Energy Convers.}, vol. 39, no. 4, pp. 2184-2201, Dec. 2024.




\bibitem{Bofan_Req}
B. Fan and X. Wang, "Equivalent Circuit Model of Grid-Forming Converters With Circular Current Limiter for Transient Stability Analysis," in \textit{IEEE Trans. Power Syst.}, vol. 37, no. 4, pp. 3141-3144, July 2022.

\bibitem{instantaneous_power}
F. Wang, J. L. Duarte and M. A. M. Hendrix, "Pliant Active and Reactive Power Control for Grid-Interactive Converters Under Unbalanced Voltage Dips," in \textit{IEEE Trans. Power Electron.}, vol. 26, no. 5, pp. 1511-1521, May 2011.

\bibitem{NERC}
North American Electric Reliability Corporation, "White Paper: Grid Forming Functional Specifications for BPS-Connected Battery Energy Storage Systems". Sep. 2023.

\bibitem{GBGF}
National Grid ESO, “Great Britain Grid Forming Best Practice Guide”. April. 2023.

\bibitem{Rui_liu}
H. Zhang, R. Liu and Y. Li, "Voltage Support Capability Analysis for Grid-Forming Inverters With Adaptive Virtual Impedance Under Asymmetrical Grid Faults," in \textit{IEEE Trans. Ind. Electron.}.















\end{thebibliography}
\end{document}